\documentclass[12pt,showpacs,preprintnumbers,superscriptaddress,amsmath,amssymb,nofootinbib]{revtex4}
\usepackage{graphicx}
\usepackage{dcolumn}
\usepackage{bm}
\usepackage{amssymb}
\usepackage{amsmath}
\usepackage{epsfig}    
\usepackage{color}
\usepackage{slashed}
\usepackage{hhline}

\def\be{\begin{equation}}
\def\ee{\end{equation}}
\newcommand{\bea}{\begin{eqnarray}}
\newcommand{\eea}{\end{eqnarray}}
\newcommand{\nn}{\nonumber}

\numberwithin{equation}{section}

\begin{document}

\title{7 keV Dark Matter as X-ray Line Signal in Radiative Neutrino Model
}
\preprint{KIAS-O14001}

\author{Seungwon Baek}
\email{swbaek@kias.re.kr}
\affiliation{School of Physics, KIAS, Seoul 130-722, Korea}
\author{Hiroshi Okada}
\email{hokada@kias.re.kr}
\affiliation{School of Physics, KIAS, Seoul 130-722, Korea}

\begin{abstract}
We study a light  dark matter  in a radiative neutrino model to explain the X-ray line signal at about $3.5$ keV 
recently reported by XMN-Newton X-ray observatory using data of various galaxy clusters and Andromeda galaxy. 
The signal requires very tiny mixing between the dark matter and an active neutrino; $\sin^2 2\theta\approx 10^{-10}$. 
It could suggest that such a light dark matter cannot contribute to the observed neutrino masses if we use the seesaw
mechanism. 
In other words, neutrino masses might come a  structure different from the dark matter.
We propose a model in which Dirac type active neutrino masses are induced at one-loop level. 
On the other hand the mixing between active neutrino and  dark matter  are generated at  two-loop level.
As a result  we can explain both the observed neutrino masses and the X-ray line signal from the
dark matter decay with  rather mild hierarchy of parameters in TeV scale.

\end{abstract}
\maketitle
\newpage

\section{Introduction}
The energy density of dark matter (DM) occupies about 27 \% of the universe. However, its nature is not known yet clearly.
Recently two groups independently reported anomalous X-ray line signal at about $3.5$ keV from the analysis of
 XMN-Newton X-ray observatory  data of various galaxy clusters and Andromeda galaxy~\cite{Bulbul:2014sua, Boyarsky:2014jta}.
In this sense, the X-ray line signal at $3.5$ keV can be explained by a 7 keV dark matter mixing 
with the active neutrino by  angle given by $\sin^2 2\theta\approx 10^{-10}$. 
This could provide a lot of implications on the nature 
of DM.  Already several works have appeared in various models~\cite{Ky:2005yq, Dinh:2006ia, Merle:2013gea, Ishida:2014dlp, Finkbeiner:2014sja, 
Higaki:2014zua, Jaeckel:2014qea,Lee:2014xua, Kong:2014gea}.
Due to its too small mixing, the DM cannot contribute to the neutrino masses directly.

In this letter, we propose a Dirac type neutrino scenario at one-loop level, introducing continuous $U(1)'$ symmetry~\cite{Lindner:2011it, Baek:2013fsa, Baek:2014awa}.
On the other hand the mixing between active neutrino and  dark matter  are generated at  two-loop level.
As a result,  such a tiny mixing can be naturally explained within TeV scale.
 
This paper is organized as follows.
In Sec.~II, we show our model for  neutrino sector and DM sector, and analyze these properties.
In Sec.~III, we summarize and conclude.


\section{The Model}
\subsection{Model setup}

\begin{table}[thbp]
\centering {\fontsize{10}{12}
\begin{tabular}{|c||c|c|c|c|c|c|}
\hline Particle & $L_L$ & $ e_{R} $  & $S_L$  &  $S_R$  & $N_R$ & $X$ 
  \\\hhline{|=#=|=|=|=|=|=|$}
$(SU(2)_L,U(1)_Y)$ & $(\bm{2},-1/2)$ & $(\bm{1},-1)$  & $(\bm{1},0)$ & $(\bm{1},0)$ & $(\bm{1},0)$  & $(\bm{1},0)$
\\\hline
$U(1)'$ & $-(3\Sigma+\Sigma')/4$ & $-(3\Sigma+\Sigma')/4$ & $(3\Sigma-\Sigma')/4$ & $(3\Sigma-\Sigma')/4$
& $(5\Sigma-\Sigma')/4$ & $3(-\Sigma+\Sigma')/4$  \\\hline
\end{tabular}%
} \caption{The particle contents and the charges for fermions.} 
\label{tab:1}
\end{table}

\begin{table}[thbp]
\centering {\fontsize{10}{12}
\begin{tabular}{|c||c|c|c|c|c|c|c|}
\hline Particle   & $\eta$  & $\Phi$  & $\chi_1$  & $\chi_2$   & $\chi_3$  & $\Sigma$   & $\Sigma'$ 
  \\\hhline{|=#=|=|=|=|=|=|=|}
$(SU(2)_L,U(1)_Y)$ & $(\bm{2},1/2)$  & $(\bm{2},1/2)$   & $(\bm{1},0)$ & $(\bm{1},0)$ & $(\bm{1},0)$ & $(\bm{1},0)$  & $(\bm{1},0)$ \\\hline
$U(1)' $  & $3\Sigma/2$ & $0$ & $-\Sigma/2$  & $-\Sigma'/2$ & $-(3\Sigma-\Sigma')/2$  & $\Sigma$   & $\Sigma'$  \\\hline
\end{tabular}%
} \caption{The particle contents and the charges for bosons. }
\label{tab:2}
\end{table}

We discuss a one-loop induced radiative neutrino model. 
The particle contents are shown in Tab.~\ref{tab:1} and Tab.~\ref{tab:2}.
Here one finds that all the terms are written in terms of the charges of $\Sigma$ and  $\Sigma'$, 
as can be seen in those tables.\footnote{We used the  same notation, $\Sigma, \Sigma'$, 
both for the fields and their $U(1)'$ charges for simplicity.}
We add three $SU(2)_L$ singlet vector like neutral fermions $S_L$ and
$S_R$, three singlet Dirac fermions $N_R$, and gauge singlet Majorana fermion $X$, which is expected to be DM.
For new bosons, we introduce $SU(2)_L$ doublet scalar $\eta$ and singlet scalars
$\chi_1$, $\chi_2$, $\chi_3$, $\Sigma$, and $\Sigma'$ to the standard model (SM). 
We suppose that  only the SM-like Higgs $\Phi$, $\Sigma$, and  
$\Sigma'$ have vacuum
expectation values (VEVs), $v/\sqrt2$, $v_\Sigma/\sqrt2$, and  $v_{\Sigma'}/\sqrt2$, respectively,
where $v=246$ GeV and ${\cal O}$(10) keV $=v_{\Sigma'}<v_{\Sigma} \ll v$ is assumed.

The continuous $U(1)'$ symmetry is imposed so as to restrict their interaction
adequately. 
Here notice that $\Sigma+\Sigma'\neq0$ to forbid the mass term of $N_RX$ that provides too large enhancement of the mixing between active neutrinos and $X$.

Moreover, we impose to generate mass term of $X$ through VEV of $\Sigma'$ for our convenience. From this 
$\Sigma=5\Sigma'/3$ is derived.

The renormalizable Lagrangian for Yukawa sector and relevant scalar potential under these assignments
are given by
\begin{eqnarray}
-\mathcal{L}_{Y}
&=&
y_\ell \bar L_L \Phi e_R +
y_{\eta} \bar L_L \eta^* S_R +
y_{\chi_1}  \bar S_L N_R \chi_1 +
y_{\chi_2}  \bar S^c_R X \chi_2 +
y_{\chi_3}  \bar S^c_R S_R \chi_3
+\rm{h.c.} \\ 
 &+&
\mu_1 \Sigma(\chi_1)^2 + \mu_2 \Sigma'(\chi_2)^2 +a(\Phi^\dag\eta)\chi_1\Sigma^* +b(\Phi^\dag\eta)\chi_2\chi_3
 +M_{S}\bar S_L S_R +\lambda \Sigma' XX 
+\rm{h.c.} 
\label{HP}
\end{eqnarray}
where the first term of $\mathcal{L}_{Y}$ can generates the charged-lepton masses, and all the couplings are assumed to be  real.
Moreover there is a mixing between $\eta$ and $\chi_1$, but $\chi_2$ and $\chi_3$ are mass eigenstate from the beginning.
The $\chi_2$ has a mass splitting between real part and imaginary part as follows:
$m_{\chi_{2_R}} = \sqrt{m_{\chi_{2_I}}^2 + 2 \sqrt{2}\mu_2 v_{\Sigma'}}$,
that plays an crucial role in generating non vanishing neutrino mass as well as mixing active neutrino and DM 
as discussed later.

After the scalar fields get vev's,  (\ref{HP}) suggests there is a remnant $Z_2$ symmetry. The odd particles under
this discrete symmetry is $L_L, e_R, N_R, X, \eta, \chi_1, \chi_2$. The lightest $Z_2$-odd particle is the lightest
neutrino and $X$ can decay into it without breaking $Z_2$ symmetry.

\subsection{Neutrino mass matrix}
The Dirac neutrino mass matrix at one-loop level as depicted in
the left hand side of Fig.~\ref{neutrino-diag} is given by~\cite{Okada:2013iba} 
\begin{align}
&(m_{\nu})_{ab}
=\sum_{\alpha=1}^3\Bigg\{
\frac{M_{S\alpha}}{64\pi^2}  (y_\eta)_{a\alpha}(y_{\chi_1})_{b\alpha}\sin2\theta_R 
I\left(\frac{m_{H_{\chi_1}}^2}{M_{S\alpha}^2},\frac{m_{H_\eta}^2}{M_{S\alpha}^2}\right) 
-[(H_{\chi_1},H_\eta,\theta_R) \to (A_{\chi_1},A_\eta,\theta_I)] \Bigg\}, \label{m_ell}
\end{align}
where
\begin{align}
I(x,y)=
\frac{-x\ln x+y\ln y +xy\ln\frac{x}{y}}{(1-x)(1-y)},\quad
\sin2\theta_R = -\frac{avv_{\Sigma}}{m^2_{H_\eta}-m^2_{H_{\chi_1}}},
\sin2\theta_I = -\frac{avv_{\Sigma}}{m^2_{A_\eta}-m^2_{A_{\chi_1}}}.
\end{align}
Here $H_i$ and $A_i$ represents real part and imaginary part of mass eigenstates.
We show a benchmark point to satisty the data of neutrino masses reported by Planck data ~\cite{Ade:2013lta}; $\sum m_{\nu}<0.933~\mathrm{eV}$,
as follows:
\begin{align}
&a\approx y_\eta y_{\chi_1}\approx1,\quad m_{H_\eta}\approx300\ {\rm GeV}, \quad m_{H_{\chi_1}}\approx 150\ {\rm GeV},
\quad M_S\approx100\ {\rm GeV}, \quad v_{\Sigma}\approx{\cal O}(1)\ {\rm MeV}, \nn\\
&(m_{H_\eta}^2-m_{A_\eta}^2)\approx (m_{H_{\chi_1}}^2-m_{A_{\chi_1}}^2)\approx\mu_1  v_{\Sigma},\quad
\sin2\theta_R\approx \sin2\theta_I\approx0.1(\ll 1).
\end{align}
Then we can obtain the neutrino mass as
\begin{align}
(m_\nu)_{ab} \approx 0.1\ {\rm eV}.
\end{align}


\subsection{Mixing between $X$ and $\nu$}

\begin{figure}[cbt]
\begin{center}
\includegraphics[scale=0.4]{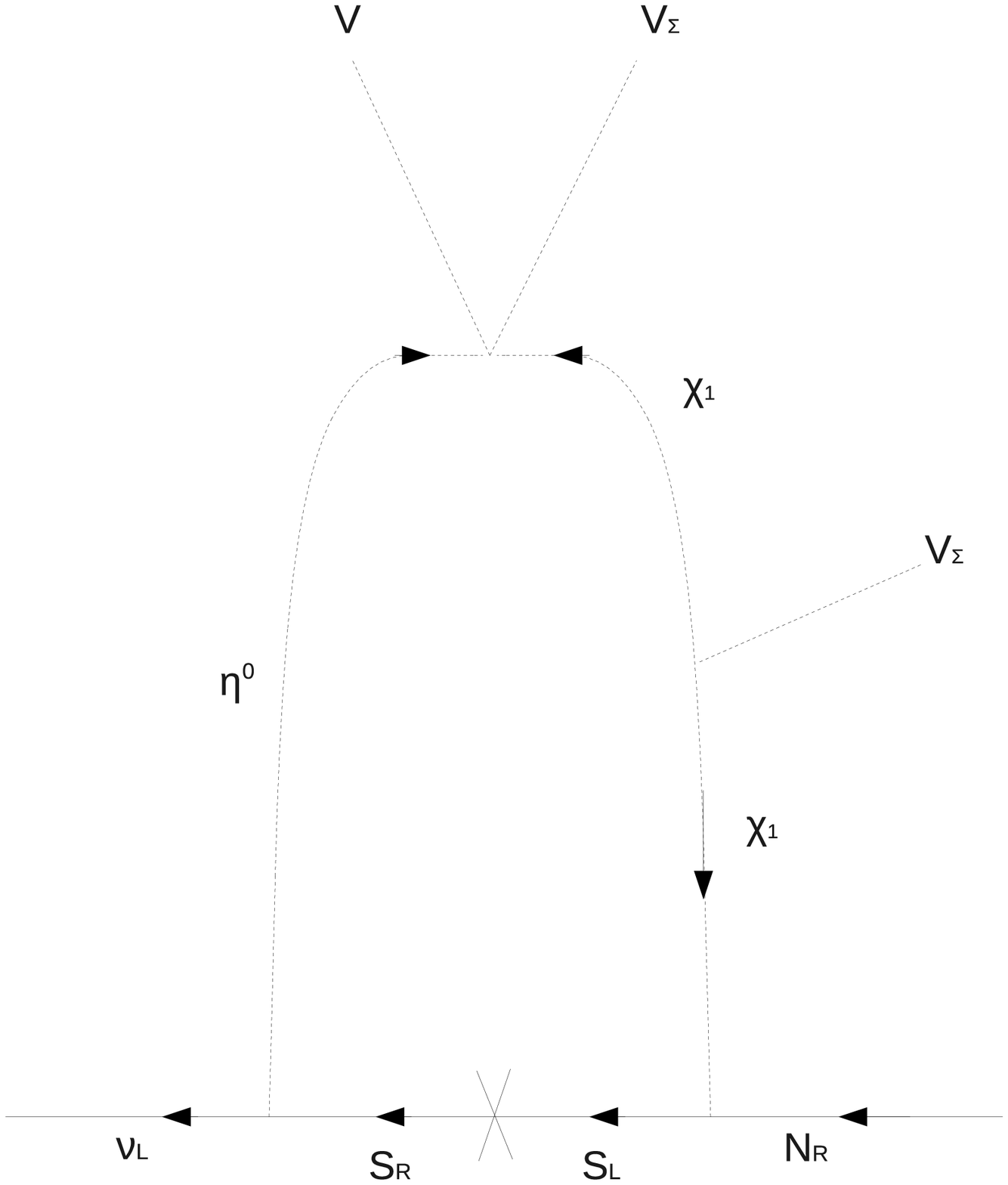}
\includegraphics[scale=0.4]{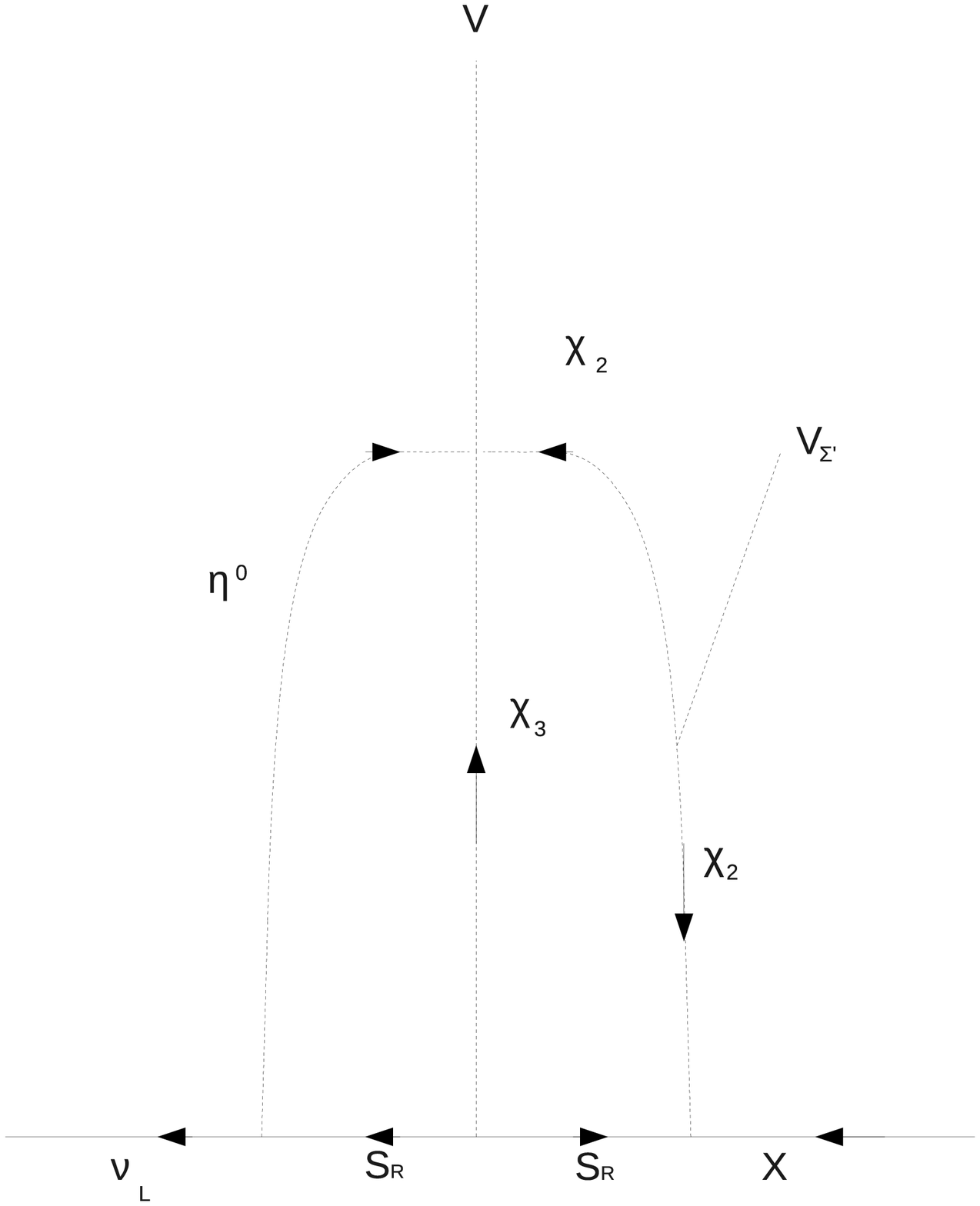}
   \caption{Diagrams for the Dirac neutrino masses (left) and radiative induced mixing between $X$ and $\nu$ (right).}
   \label{neutrino-diag}
\end{center}
\end{figure}
\begin{figure}[tbc]
\begin{center}
\includegraphics[scale=0.8]{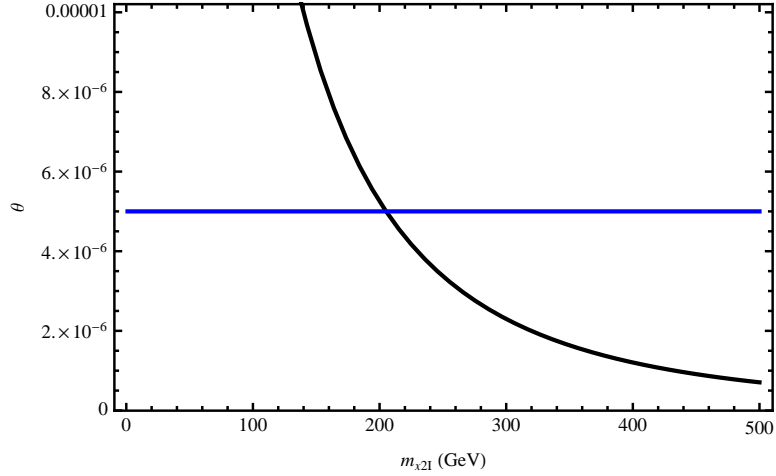}
   \caption{Order estimation for the mixing angle $\theta$ between active neutrino and DM as a 
function of the $\chi_{2I}$ mass.
Here we fixed $\theta_R\approx\theta_I \ll 1$, $ b(y_{\eta}y^*_{\chi_3}y_{\chi_2})\approx 1.0$, $M_{S}=$100 GeV, $m_{H_\eta}=$300 GeV, $m_{\chi_3}=$200 GeV,
   $\mu_2=$1 TeV, and $v_{\Sigma'}=$10 keV. The black line represents 
$\theta$ given in Eq~(\ref{eq:mixing}).  
   The blue line, $\theta\approx5\times10^{-6}$, is the expected 
mixing angle to explain X-ray line signal reported by~\cite{Ade:2013lta}.
   }
   \label{mix-plot}
\end{center}
\end{figure}

The mixing mass term between $X$ and $\nu$ are obtained at two-loop level as depicted in
the left hand side of Fig.~\ref{neutrino-diag}, and its form is given by 
\begin{align}
(m_{\nu-X})_{ab}
&\approx
\sum_{c=1}^{3} \sum_{d=1}^{3}
\left[ \frac{v(y_\eta)_{ac}(M_S)_{c}(y_{\chi_3})^*_{cd}(M_S)_d(y_{\chi_2})_{db}}{8\sqrt2(4\pi)^4(M_{Sc})^2}
\right]
\nn\\
&\times 
\left[ F\left(\frac{m_{H_{\eta}}^2}{M_{Sc}^2}, \frac{m_{\chi_3}^2}{M_{Sc}^2},\frac{m_{\chi_{2R}}^2}{M_{Sc}^2}\right) 
-
F\left(\frac{m_{H_{\eta}}^2}{M_{Sc}^2}, \frac{m_{\chi_3}^2}{M_{Sc}^2},\frac{m_{\chi_{2I}}^2}{M_{Sc}^2}\right) 
\right],
\label{eq:neutrinomass}
\end{align}
where we assume $\theta_{R{I}} \ll 1$, and the loop function $F$ is computed by
\begin{align}
F\left(X_1,X_2,X_3\right) 
&=
\int_0^1dx\int_0^1dy \int_0^1dz\delta(x+y+z-1)\int_0^1d\alpha\int_0^1d\beta \int_0^1d\gamma\delta(\alpha+\beta+\gamma-1)
\nn\\&\times 
\frac{1}{(z^2-z)(\alpha+\beta X_1-\gamma \Delta)},\
\Delta=\frac{x\frac{M_{Sd}^2}{M_{Sc}^2}+yX_2+zX_3}{(z^2-z)}.
\end{align}
The mixing between active neutrino and DM is given as
\begin{align}
\theta\equiv \frac{m_{\nu-X}}{m_X}\approx5\times10^{-6}\label{eq:mixing},
\end{align}
where $m_X=\lambda v_{\Sigma'}/\sqrt2(\approx$7.5 keV) is the mass of DM, and $5\times10^{-6}$ is 
an expected mixing angle from the X-ray experiment~\cite{Bulbul:2014sua,Boyarsky:2014jta}. 
In Fig.~\ref{mix-plot} we show the mixing $\theta$ as a function of the imaginary mass of $\chi_2$, $m_{\chi_{2I}}$.  
The figure shows that we can get the required mixing angle near $m_{\chi_{2I}}=$200 GeV.
We fixed $\theta_R\approx\theta_I<<1$, $b(y_{\eta}y^*_{\chi_3}y_{\chi_2})\approx 1.0$, $M_{S}=$100 GeV, $m_{H_\eta}=$300 GeV, $m_{\chi_3}=$200 GeV,
$\mu_2=$1 TeV, and $v_{\Sigma'}=$10 keV for the figure. 


\section{Summary and Conlusion}
We have  shown that a light DM   in a radiative model can explain the X-ray line signal 
from the XMN-Newton X-ray observatory  data of various galaxy clusters and Andromeda galaxy . 
We can accommodate both the observed neutrino masses and  the DM  ($m_{X}=7$ keV) mixing weakly with 
active neutrino ($\sin^22\theta\approx10^{-10}$) for the signal by a benchmark point  with a rather mild hierarchy 
at TeV scale; 
\begin{align}
&a\approx y_\eta y_{\chi_1}\approx1,\quad m_{H_\eta}\approx300\ {\rm GeV}, \quad m_{H_{\chi_1}}\approx 150\ {\rm GeV},
\quad M_S\approx100\ {\rm GeV}, \quad v_{\Sigma}\approx{\cal O}(1)\ {\rm MeV},  \nn\\
&(m_{H_\eta}^2-m_{A_\eta}^2)\approx (m_{H_{\chi_1}}^2-m_{A_{\chi_1}}^2)\approx\mu_1  v_{\Sigma},\quad
\sin2\theta_R\approx \sin2\theta_I\approx0.087(<<1),\quad v_{\Sigma'}=10 \ {\rm keV},\nn\\
&m_{\chi_{2I}}\approx 200\ {\rm GeV},\quad b(y_{\eta}y^*_{\chi_3}y_{\chi_2})\approx 1.0, \quad m_{H_\eta}=300\ {\rm GeV}, m_{\chi_3}=200\ {\rm GeV}, \quad \mu_2=1\ {\rm TeV}.
\end{align}
The relic density of DM  can be thermally obtained through an additional gauged boson, if our $U(1)'$ is localized~\cite{Kusenko:2009up, Ishida:2013mva, Khalil:2008kp},
or it can be also obtained  non-thermally ~\cite{Petraki:2007gq}.



\end{document}